\documentclass[12pt]{nature}
\usepackage{graphicx,color,multirow,bm,braket,amsmath,soul,epstopdf,lineno}
\usepackage[final]{pdfpages}

\def\bk{{\bf k}}
\def\bq{{\bf q}}
\def\ve{\varepsilon}

\begin{document}

\title{Origin of the crossover from polarons to Fermi liquids in transition metal oxides}
\author{Carla Verdi, Fabio Caruso, Feliciano Giustino}

\maketitle 

\vspace{0.2cm} 
\begin{affiliations} 
\item[] Department of Materials, University of Oxford, Parks Road, Oxford OX1 3PH, United Kingdom.
\\ \small \normalfont Correspondence and requests for materials should be addressed to 
F.G. (email: feliciano.giustino@materials.ox.ac.uk).
\end{affiliations} 
 
\begin{abstract} 
Transition metal oxides (TMOs) host a wealth of exotic phenomena ranging from charge, orbital, and 
magnetic order to nontrivial topological phases and superconductivity. In order to translate these 
unique materials properties into novel device functionalities, TMOs must be doped. However, 
the nature of carriers in doped oxides and their conduction mechanism at the atomic scale 
remain unclear. Recent angle-resolved photoelectron spectroscopy (ARPES) investigations provided new 
insight into these questions, revealing that the carriers of prototypical metal oxides undergo a 
transition from a polaronic liquid to a Fermi liquid regime with increasing doping. 
Here, by performing \textit{ab initio} many-body calculations of the ARPES spectra of 
TiO$_{\textbf{2}}$, we show that this transition originates from non-adiabatic polar electron-phonon 
coupling, and occurs when the frequency of plasma oscillations exceeds that of longitudinal-optical 
phonons. This finding suggests that a universal mechanism may underlie polaron formation 
in TMOs, and provides a new paradigm for engineering emergent properties in quantum matter.
\end{abstract}
\vspace{0.3cm} 

Elucidating the nature of charge carriers in doped transition metal oxides is key to 
understanding the mechanism of electrical conduction in these multifunctional materials. In 
conducting oxides the infrared-active vibrations can couple strongly to electrons, leading to the 
formation of polarons\cite{Devreese2009}. Polarons are electrons dressed by a phonon cloud\cite{Mahan}, 
and represent a paradigmatic example of emergent state in condensed matter.
Depending on their mass and size, polarons exhibit widely different conduction 
mechanisms, from band-like transport to thermally-activated hopping transport\cite{Mott1969, Ziman}. 
Despite being central to the science and technology of oxides, little is known about the 
properties of polaronic states.

The interest in electron-phonon coupling and polaronic quasiparticles in TMOs has been reinvigorated 
by recent ARPES experiments\cite{Moser2013,Chen2015,Cancellieri2016,Baumberger2016,Yukawa2016}. 
The signature of polaronic behavior in ARPES spectra is the appearance of satellites below the 
conduction band, at integer multiples of the optical phonon energy. This is reported 
in Fig.~\ref{fig1}a-b for the paradigmatic case doped anatase TiO$_2$\cite{Moser2013}. These 
pioneering measurements showed that, by increasing the carrier concentration, polaronic satellites 
gradually evolve into the photoemission kinks observed in metals and superconductors\cite{Damascelli2003}, 
see Fig.~\ref{fig1}c. It was proposed that this crossover reflects the evolution of charge carriers from 
polarons to a Fermi liquid\cite{Moser2013,Baumberger2016}. In order to clarify the origin of this transition 
without making any {\it a priori} assumption about the underlying mechanism, first principles 
calculations are urgently called for. However, the investigation of polaronic features 
in ARPES spectra from first principles and their evolution with doping is exceptionally challenging 
and has never been reported before.

In the following we focus on the prototypical example of anatase TiO$_2$. On top of its well-known 
applications in solar energy harvesting\cite{Hardin2012, Snaith2014} and superhydrophilic 
technology\cite{Fujishima1972, Fujishima2000}, this material is also being investigated in the quest 
for transparent conducting oxides based on non-toxic and Earth-abundant elements\cite{Furubayashi2005,
Ellmer2012}. Despite its pivotal role in a broad range of technologies, the nature of the charge carriers 
in anatase is still controversial\cite{DeAngelis2014}. Here we address these issues by calculating ARPES 
spectra and polaron wavefunctions entirely from first principles. We develop a novel theoretical and 
computational framework that allows us to investigate polarons and Fermi liquid quasiparticles on the 
same footing, and without resorting to any empirical parameters. Using this new approach, we show how 
the interplay between the dynamical screening of the electron plasma and the Fr\"ohlich electron-phonon 
coupling is responsible for the transition between polaronic and Fermi liquid states. We propose that 
the mechanism identified in this work may be universal, and also applies to 
other oxides such as SrTiO$_3$ and ZnO. 

\vspace{10pt}

\section*{Results} 

\section*{\small Angle-resolved photoemission spectra}

\begin{figure}[t!]
  \begin{center} 
  \includegraphics[width=\textwidth]{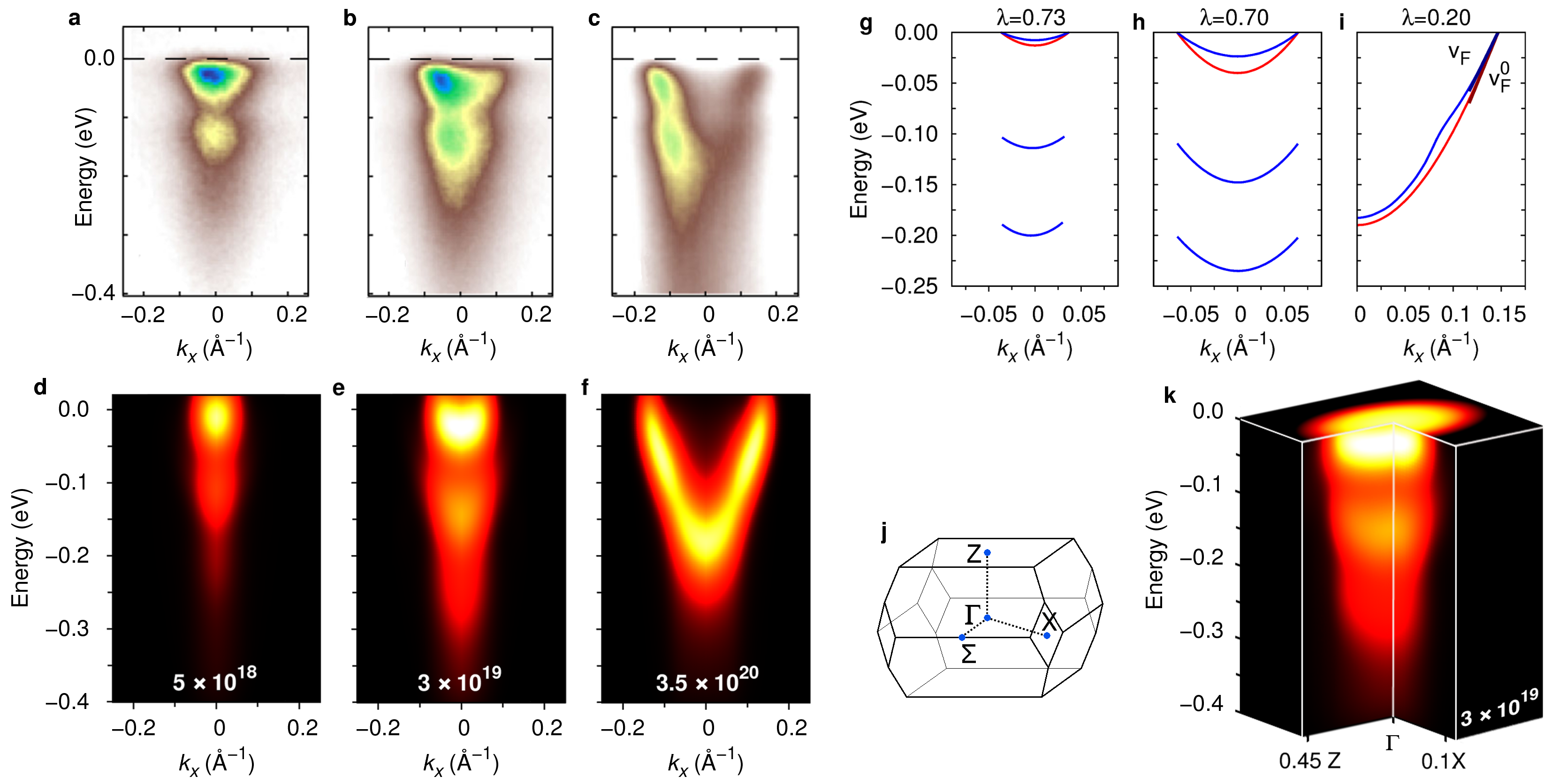}
  \end{center} 
  \caption{\label{fig1} \small
  \textbf{$\bm|\,$ \textit{Ab initio} ARPES spectra of \textit{n}-doped anatase TiO$_{\textbf{2}}$.}
  (\textbf{a}-\textbf{c}) ARPES spectra of anatase TiO$_2$ measured by Moser {\it et al.}\cite{Moser2013}.
  The measurements were taken at 20~K on samples with doping concentrations $5\times10^{18}$~cm$^{-3}$ 
  (\textbf{a}), $3\times10^{19}$~cm$^{-3}$ (\textbf{b}), and $3.5\times10^{20}$~cm$^{-3}$ (\textbf{c}). The
  zero of the energy is set to the Fermi level. The electron momentum $k_x$ is along the 
  $\Gamma\Sigma$ line of the anatase Brillouin zone (see panel {\bf j}). Reproduced with permission from 
  ref.~\citenum{Moser2013}, copyright 2013 American Physical Society. 
  (\textbf{d}-\textbf{f}) Calculated spectral function of anatase TiO$_2$, for the same electron momenta 
  and nominal doping levels as in {\bf a}-{\bf c} (indicated in each panel in units of cm$^{-3}$).
  Gaussian masks of widths 25~meV and 0.015~\AA$^{-1}$ were applied to account for the experimental
  resolution\cite{Moser2013}. (\textbf{g}-\textbf{i}) Band structures extracted from the calculated 
  spectral functions in \textbf{d}-\textbf{f}. The bare bands are in red, the bands including electron-phonon 
  interactions are in blue. The calculated mass enhancement parameter $\lambda$ is indicated in each panel. 
  (\textbf{j}) Brillouin zone and high-symmetry lines of anatase TiO$_2$.
  (\textbf{k}) Calculated ARPES spectrum for a doping concentration of $3\times10^{19}$~cm$^{-3}$, 
  showing the anisotropy of the electron dispersions along $\Gamma X$ (basal plane of the tetragonal 
  lattice, see Supplementary Fig.~1) and $\Gamma Z$ ($c$-axis). }
  \end{figure}

Our calculated ARPES spectra are shown in Fig.~\ref{fig1}\mbox{d-f}, for the same doping levels as in the 
measurements of ref.~\citenum{Moser2013}, reproduced in Fig.~\ref{fig1}a-c. These maps show the bottom 
of the conduction band of \mbox{$n$-doped} anatase TiO$_2$, for three doping levels in the range 
$10^{18}$ to $10^{20}$~cm$^{-3}$. All the spectra exhibit a bright parabolic band, whose
size increases with doping. This reflects the rise of the Fermi energy inside the conduction band 
as the electron density increases. Besides this bright feature, panels a-b (experiments) and d-e
(calculations) show each a pair of satellites, a bright one at a binding energy around 0.1~eV, and
a dim one near 0.25~eV. These features are identified as polaronic effects\cite{Moser2013}. Moving on 
to higher doping in panels c and f, the satellites disappear and are replaced by band structure kinks
near 0.1~eV. Overall, our calculated ARPES spectra are in remarkable agreement with the experiments 
of ref.~\citenum{Moser2013}. In order to achieve this unprecedented level of precision without any 
adjustable parameters, we developed an innovative computational framework.

In our calculations the photoelectron intensity maps are obtained using the single-particle spectral 
function $A(\bk,\omega)$, where $\hbar\bk$ and $\hbar\omega$ are the electron momentum and binding
energy, respectively, and $\hbar$ is the reduced Planck constant. The spectral function is calculated 
using the state-of-the-art cumulant expansion, which has been employed recently to successfully 
describe plasmon satellites\cite{Aryasetiawan1996,Guzzo2011,Lischner2013,Caruso2015}. 
The cumulant expansion can naturally be applied to investigate the spectral properties of a polaronic 
system since the theory stems from the exact solution of the Fr\"ohlich electron-boson coupling 
Hamiltonian\cite{Langreth1970,Rehr2014}. In this formalism the spectral function is expressed 
as\cite{Hedin1980}:
  \begin{equation} \label{eq.spec.cum}
  A(\bk,\omega) = \frac{1}{2\pi}\,{\sum}_n {\rm Re}\! \int_{-\infty}^{+\infty}\! \!dt\,
  e^{i(\omega-\ve_{n\bk}/\hbar) t}\,e^{C_{n\bk}(t)},
  \end{equation}
where $n$ is the electron band index, $\ve_{n\bk}$ is the electron eigenvalue in absence of many-body 
interactions, and $t$ is the time variable. The quantity $C_{n\bk}$ is the so-called `cumulant'
function, and can be calculated by using the standard electron-phonon self-energy $\Sigma_{n\bk}$ as a 
seed\cite{Gumhalter2016,GiustinoRMP} (see Methods and Supplementary Note~1):
  \begin{align} \label{Sigma} 
  \Sigma_{n\bk}(\omega)&=\frac{1}{\hbar}\sum_{m\nu}\int \!\frac{d\bq}{\Omega_{\rm BZ}} 
  |g_{mn\nu}(\bk,\bq)|^2 
  \times \left[\frac{n_{\bq\nu}+f_{m\bk+\bq}}{\omega-\ve_{m\bk+\bq}/\hbar
  +\omega_{\bq\nu}-i\eta} +\frac{n_{\bq\nu}+1-f_{m\bk+\bq}}{\omega-\ve_{m\bk+\bq}/\hbar \!-
  \omega_{\bq\nu}-i\eta} \right] .
  \end{align}
Here $g_{mn\nu}(\bk,\bq)$ is the electron-phonon vertex, and describes the probability amplitude for 
an electron in the initial state $|n\bk \rangle$ to be scattered into the final state $|{m\bk+\bq}\rangle$ 
by a phonon with momentum $\hbar\bq$ and energy $\hbar\omega_{\bq\nu}$ in the branch $\nu\,$\cite{GiustinoRMP}. 
The terms $n_{\bq\nu}$ and $f_{m\bk+\bq}$ denote the Bose-Einstein and 
Fermi-Dirac occupations, respectively; $\Omega_{\rm BZ}$ is the Brillouin zone volume, and $\eta$ 
a positive infinitesimal. The numerical evaluation of Eqs.~\eqref{eq.spec.cum}-\eqref{Sigma} is very 
challenging owing to the singular nature of the Fr\"ohlich interaction at long wavelengths\cite{Frohlich1954}. 
To overcome this challenge we use the Wannier function technique of refs.~\citenum{EPW2007,Verdi2015}, 
as implemented in the \texttt{EPW} code\cite{EPW2016}. 

The strength of the electron-phonon interaction is most commonly expressed in terms of a single 
parameter $\lambda$, which describes the enhancement of the electron mass from the band effective 
mass $m_{\rm b}$ to the polaron mass $m^*$ via $m^*=m_{\rm b}(1+\lambda)$, as well as 
the renormalization of the Fermi velocity\cite{Grimvall}. 
To determine this parameter we first extract the band structures underlying \mbox{Fig.~\ref{fig1}d-f}, by 
tracking the maxima in the energy distribution curves. This procedure yields a set of parabolic bands and 
their satellites in \mbox{Fig.~\ref{fig1}g-h}, and a distorted parabola in \mbox{Fig.~\ref{fig1}i} (blue 
curves). For comparison we also report the electronic bands in absence of electron-phonon interactions 
(red curves). For carrier concentrations of $5\times10^{18}$ and $3\times10^{19}$~cm$^{-3}$, polaron 
satellites are clearly visible in \mbox{Fig.~\ref{fig1}g-h}, whereas at $3.5\times10^{20}$~cm$^{-3}$ 
we see a band structure kink but no satellites. 
From these band structures we obtain the mass enhancement parameter as the ratio 
between the Fermi velocity of the bare band, $v^0_{\rm F}$, and that of the dressed band, $v_{\rm F}$, 
as indicated in Fig.~\ref{fig1}i: $\lambda_{\bf k}=v^0_{\bf k\rm F}/v_{\bf k\rm F} - 1$~\cite{Grimvall}, 
where we explicitly included the dependence on the wavevector at the Fermi surface. 
As we move from the lowest to the highest doping level we obtain $\lambda_{\bf k}=0.73$, 0.70, and 0.20,
respectively. Our calculated value at intermediate doping is in excellent agreement with the mass enhancement 
determined in experiments, $\lambda=0.7$~\cite{Moser2013}. 

In Fig.~\ref{fig1}k we show the spectrum calculated at intermediate doping for electron momenta 
along $\Gamma X$ and $\Gamma Z$, as well as the Fermi surface cut. Here we see that the Fermi 
surface pocket is elongated along the $\Gamma Z$ direction. This elongation reflects the anisotropic 
character of the band effective masses, which we calculate to be $m_{\rm b}^\perp =0.40\,m_{\rm e}$ and 
$m_{\rm b}^\parallel =4.03\,m_{\rm e}$ along $\Gamma$X and $\Gamma$Z, respectively ($m_{\rm e}$ is the 
electron mass). Surprisingly this anisotropy is not reflected in the electron-phonon coupling strength: 
our calculations indicate that the mass enhancement $\lambda_{\bf k}$ varies by less than 10\% along 
the [100], [110], and [001] directions, leading to an average value of $\lambda=0.73$, 0.70 and 0.19 
for the three doping levels considered (see also Supplementary Note~2). These results are in good 
agreement with resonant inelastic X-ray scattering experiments on anatase TiO$_2$\cite{Moser2015}. 

\section*{\small Origin of satellites and kinks}

\begin{figure}[p] 
 \begin{center} 
 \includegraphics[width=\textwidth]{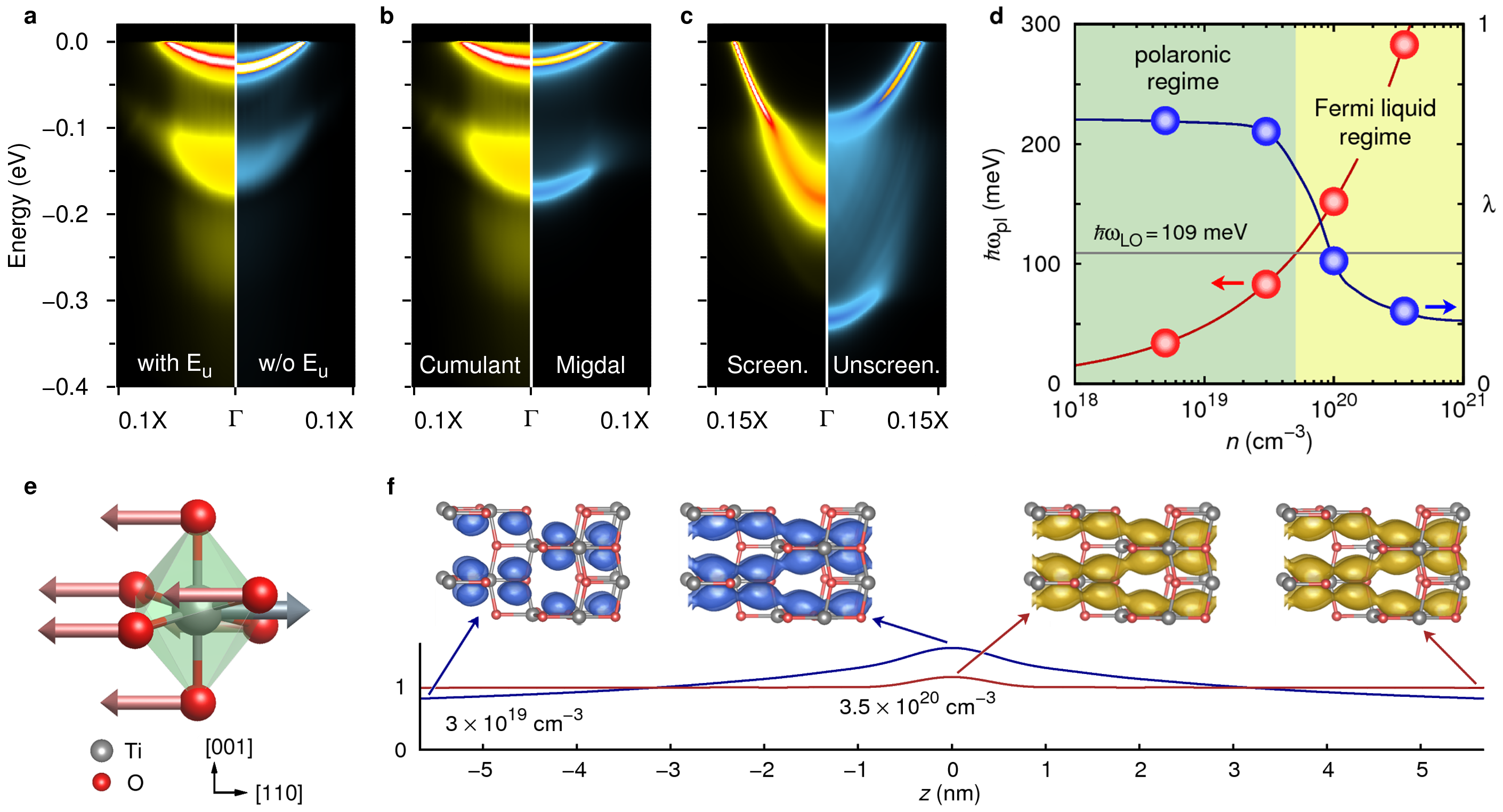}
 \end{center}
 \caption{\label{fig2} \small
 \textbf{$\bm|\,$ Origin of satellites and kinks, and polaron wavefunctions in anatase TiO$_\textbf{2}$.}
 (\textbf{a}) Effect of high-energy phonons: we compare the spectral function calculated  by taking into 
 account all vibrational modes (left half-panel, for $3\times10^{19}$~cm$^{-3}$) and a calculation where 
 all phonons with energy above 100~meV have been eliminated (right half-panel). 
 (\textbf{b}) Effect of electron-phonon correlations beyond the one-shot Migdal approximation: 
 the left half-panel shows a complete calculation using the cumulant expansion method, while in the right 
 half-panel the spectral function is calculated within the one-shot Migdal approximation. 
 The doping level is the same as in \textbf{a}.
 (\textbf{c}) Effect of dynamical screening on the spectral function:
 the left half-panel shows the spectral function calculated for the highest doping level
 ($3.5\times10^{20}$~cm$^{-3}$) by taking into account the screening of the electron-phonon 
 interaction by carriers. In the right half-panel this effect is turned off, and as a result the 
 electron-phonon coupling is artificially enhanced. For clarity these spectra were not convoluted 
 with Gaussian masks as in Fig.~\ref{fig1}d-f. (\textbf{d}) Identification of polaronic 
 region and Fermi liquid region in $n$-doped anatase TiO$_2$: the red spheres represent the plasmon 
 energy at each doping level, the horizontal line is the energy of the LO $E_u$ phonon. The 
 electron-phonon coupling strength $\lambda$ is given by the blue spheres. The lines are guides 
 to the eye. (\textbf{e}) Ball-and-stick representation of the LO $E_u$ phonon, showing 
 for clarity only one of the TiO$_6$ octahedra. 
 (\textbf{f}) Square moduli of the polaron wavefunctions near the origin and further away 
 from the origin, in the polaronic region (blue, $3\times10^{19}$~cm$^{-3}$) and in the fermi liquid 
 region (gold, $3.5\times10^{20}$~cm$^{-3}$). The corresponding envelope functions are shown 
 in the bottom half of the figure as the blue and red curves, respectively. These 
 wavefunctions are extended over all three Cartesian directions, but are only shown along the $c$-axis 
 for clarity. }
 \end{figure}

Having established that our calculations can accurately reproduce experimental spectra without 
adjustable parameters, we now proceed to identify the mechanisms that drive the formation of 
polaron satellites by selectively turning off individual components of the calculations. 
The energy separations between the quasiparticle bands and the first satellite in Fig.~\ref{fig1}d 
and Fig.~\ref{fig1}e are 106~meV and 124~meV, respectively. In the case of Fig.~\ref{fig1}f 
the kink appears at a binding energy of 100~meV. These energy scales are compatible with a 
Fr\"ohlich-type coupling to the longitudinal-optical (LO) $E_u$ phonon at 109~meV (see phonon 
dispersions in Supplementary Fig.~1). This vibrational mode corresponds to the stretching of 
the Ti-O bonds in the $ab$ plane, as shown in Fig.~\ref{fig2}e. Another candidate bosonic mode is 
the $c$-axis $A_{2u}$ phonon at 88~meV\cite{Moser2015}, however the energy of this mode appears too 
small to account for the satellites and kinks in Fig.~\ref{fig1}. In order to quantify the importance 
of the $E_u$ phonon, in Fig.\ref{fig2}a we compare two calculations: the complete spectrum at intermediate 
doping (left) and a calculation where the coupling to modes with energy above 100~meV is artificially 
suppressed (right). We see that, upon removing high-energy phonons, the intensity of the first satellite 
decreases and the second satellite disappears. The effective mass is also visibly lower, in fact the 
analysis of the mass enhancement parameter yields $\lambda=0.3$, to be compared to the total coupling 
$\lambda=0.7$. It follows that the $E_u$ phonon contributes $\sim60\%$ of the total coupling, hence it 
represents the primary mechanism behind the satellites.

In Fig~\ref{fig2}b we test the importance of many-body correlations beyond the one-shot 
Migdal approximation. On the left we show the spectrum at intermediate doping, obtained 
from the cumulant expansion; on the right we show the result of a calculation within the one-shot 
Migdal approximation (see Methods). The Migdal approximation is obtained from the
complete electron-phonon self-energy by neglecting the three-point vertex\cite{Mahan}. In
this approximation the electron-phonon self-energy contains only non-crossing diagrams. 
The additional approximation adopted in first principles calculations is to replace
the fully renormalized  electron and phonon propagators by those evaluated within
density functional theory\cite{GiustinoRMP}. We name this choice the `one-shot' Migdal
approximation to emphasize the lack of self-consistency in the electron propagator. 
From Fig.~\ref{fig2}b we notice that the one-shot Migdal approximation fails twice: firstly, the separation 
between the quasiparticle band and the first satellite is too large as compared to experiment (151~meV instead 
of $\sim100$~meV); secondly, the dim satellite around 0.25~eV is completely missing. This test highlights 
the crucial role of high-order electron-phonon correlations in the description of polarons in conducting 
oxides. It would be interesting to see whether the deficiencies of the one-shot Migdal approximation 
could be avoided by performing a fully self-consistent Migdal calculation,\cite{Marsiglio1990} and to 
establish the importance of crossing diagrams which are included in the cumulant expansion; however 
this test is currently out of reach.

We now move to discuss the origin of the crossover between satellites and kinks in the spectra as a 
function of doping. The Fr\"ohlich theory of polarons\cite{Frohlich1954} considers a single electron 
added to a polar insulator. This description does not take into account that, as more electrons are 
added to the system, the polar electron-phonon coupling is weakened by the electronic screening of the 
charge carriers. Following ref.~\citenum{Mahan}, we treat this issue by screening the electron-phonon 
vertex by the frequency-dependent Lindhard dielectric function with the calculated effective mass and 
dielectric permittivity of anatase TiO$_2$ (see also Methods). For completeness we show in Supplementary 
Fig.~2 the dielectric screening as a function of carrier density and how the Fr\"ohlich electron-phonon 
vertex is influenced by doping. In order to illustrate the importance of carrier screening, here in 
Fig.~\ref{fig2}c we compare two scenarios: the spectrum calculated at high doping by accounting for 
electronic screening (left); the same system, but this time ignoring the screening of the electron-phonon 
coupling by doped carriers (right). The result is striking: in the absence of screening one obtains a 
sharp polaron satellite, in stark disagreement with experimental evidence. This comparison indicates 
that a correct description of the electronic screening is absolutely crucial to capture the photoemission 
kink at high doping. On the contrary, when we repeat the comparison of Fig.~\ref{fig2}c for the cases of 
low and intermediate doping, we find a completely different picture: at these doping levels the electronic 
screening does not play any significant role. 

These conflicting observations can be rationalized by inspecting the timescales of lattice vibrations 
and electronic screening. The $E_u$ phonon vibrates with a period $T_{\rm ph}=38$~fs. The characteristic 
response time of the electronic screening is set by the plasma frequency of doped carriers, $T_{\rm el} = 
2\pi/\omega_{\rm p}$. In the case of $n$-doped TiO$_2$ the electrons occupy a singly-degenerate parabolic 
band minimum, therefore $\omega_{\rm p}= (n e^2/\epsilon_0 \epsilon_\infty m_{\rm b})^{1/2}$, 
where $n$ is the electron density, $\epsilon_0$ the permittivity of vacuum, and 
$\epsilon_\infty$ the high-frequency dielectric constant of TiO$_2$\cite{Mahan}. For the electron 
densities considered in Fig.~\ref{fig1}a-c we calculate $T_{\rm el} = 122$~fs, 50~fs, and 15~fs, 
respectively. From these values we deduce that, at the lowest doping, the carriers are too slow to 
screen the long-range electric field generated by the oscillation of the $E_u$ phonon (122~fs vs.\ 
38~fs). In this case the screening is ineffective and the Fr\"ohlich interaction dominates the spectrum. 
On the contrary, at the highest doping the electrons oscillate faster than the LO phonon (15~fs vs.\ 
38~fs). In this case the electronic screening is almost complete and the Fr\"ohlich coupling is largely 
suppressed. In this regime the strength of the kink depends critically 
on the carrier concentration; with increasing doping the coupling to the LO phonons 
is gradually suppressed, and the ARPES spectrum is 
dominated by the weaker coupling of carriers to non-polar phonons
\cite{Baumberger2016}. These considerations are summarized in Fig.~\ref{fig2}d, where we compare the 
energy of the $E_u$ phonon with the plasma energy $\hbar\omega_{\rm p}$ of the carriers, and we monitor 
the evolution of the coupling strength with doping. We can identify two regions: a  polaronic regime, 
corresponding to the situation $\omega_{\rm p}<\omega_{\rm ph}$, and a  Fermi liquid regime, corresponding 
to $\omega_{\rm p} > \omega_{\rm ph}$. In the polaronic region the electronic screening is ineffective, 
we see satellites in the spectra, and the electron-phonon mass enhancement is not sensitive to the doping 
level. In the Fermi liquid region the Fr\"ohlich coupling is strongly suppressed, polaron satellites are 
gradually replaced by photoemission kinks, and the coupling strength decreases. 
To further validate this trend, we performed additional calculations for a doping concentration of 
$1\times10^{20}$~cm$^{-3}$ in the transition region. We found that one satellite is still present in 
the spectrum (Supplementary Fig.~3). A careful investigation of the mass renormalization parameter 
yields \mbox{$\lambda=0.34$}, thus confirming that the electron-phonon coupling is weakened by the electronic 
screening. The present analysis reveals that the origin of the crossover from polarons to a Fermi liquid 
in the ARPES spectra of doped TiO$_2$ is to be found in a novel form of electron-phonon coupling, which 
we refer to as a `non-adiabatic Fr\"ohlich interaction'. The non-adiabatic treatment of the electronic 
screening is especially important to correctly capture the response of the electrons at low doping.

Given the qualitative change in the band structures at the crossover from the polaronic to the Fermi 
liquid regime, it is natural to ask how this evolution is reflected in the wavefunctions. In order to 
explore this aspect we calculate the polaron wavefunctions by generalizing the perturbation theory of 
ref.~\citenum{Mahan} to {\it ab initio} calculations. The top of Fig.~\ref{fig2}f shows the 
square moduli of the polaron wavefunctions at the bottom of the conduction band near
the maximum of the polaron wavefunction and a few unit cells away: on the left the intermediate doping 
level, which is inside the polaronic region of Fig.~\ref{fig2}d; on the right the high 
doping level, in the Fermi liquid region. The bottom half of Fig.~\ref{fig2}f 
shows the corresponding
envelope functions. Here we see that the wavefunction of an electron in the Fermi liquid region 
is essentially akin to a periodic Bloch function. On the contrary, the wavefunction of an electron 
in the polaronic region shows spatial localization. To quantify the `size' of the polaron in this 
latter case we define a polaron radius $r_{\rm p}$ using the half width at half maximum. From 
Fig.~\ref{fig2}f  we obtain $r_{\rm p}=5.7$~nm. This result indicates that, despite 
the qualitative changes in the ARPES spectra, we are in the presence of large polarons throughout 
the entire doping range, and that polarons in TiO$_2$ are considerably more delocalized than
previously thought\cite{Moser2013}. Since thermally-activated hopping transport corresponds 
to $r_{\rm p}$ of the order of the lattice constant\cite{Emin}, our finding supports the 
notion that electrical conduction in anatase TiO$_2$ takes place via standard band-like transport. 
To avoid ambiguities we emphasize that the present result refers to intrinsic
mobile polarons, not to localized electronic defect states such as those associated with O vacancies
and which are not mobile\cite{Setvin2014}.

\vspace{10pt}

\section*{Discussion}

The non-adiabatic Fr\"ohlich mechanism identified here is simple enough that it is likely to play
a role in many other conducting oxides. In order to test this hypothesis we estimate the critical 
density for the crossover in doped SrTiO$_3$\cite{Baumberger2016} and doped ZnO\cite{Yukawa2016}. 
In SrTiO$_3$ the plasma energy for an electron concentration of $10^{20}$~cm$^{-3}$ is 
$\hbar\omega_{\rm p}\sim64$~meV\cite{Gervais1993}, therefore to match the 100~meV LO phonon of 
SrTiO$_3$ one would need a doping level of $2.6\times10^{20}$~cm$^{-3}$. In the experiments 
of ref.~\citenum{Baumberger2016} the carriers are confined in a thin surface layer corresponding 
to approximately 3~unit cells, therefore we estimate the critical carrier density for the two-dimensional 
electron liquid in the range $3\times10^{13}$~cm$^{-2}$. This value is remarkably close to the 
critical density determined experimentally, $4\times10^{13}$~cm$^{-2}$~\cite{Baumberger2016}. 
More accurate calculations will need to take into account the two-dimensional screening at the 
surface of SrTiO$_3$ and its quantum paraelectric nature. 
A similar analysis can be carried out for ZnO\cite{Yukawa2016}. In this case the energy of the 
highest LO phonon is 72~meV, while the plasma energy for a surface electron concentration of 
$7.5\times10^{13}$~cm$^{-2}$ is 320~meV\cite{Goldstein1982}. The transition would then appear 
around $3.8\times10^{12}$~cm$^{-2}$, slightly below the density reported in ref.~\citenum{Yukawa2016}. 
In agreement with our estimate, the spectra of ref.~\citenum{Yukawa2016} exhibit a behavior 
which is intermediate between kinks and satellites. 

In summary, our findings indicate that the electron-phonon coupling in TiO$_2$ is more complex than 
previously thought, and that the non-adiabatic Fr\"ohlich coupling could be the unifying mechanism 
behind the transition from polarons to Fermi liquids in conducting oxides; explicit
{\it ab initio} calculations will be required to confirm this point. 
Looking further ahead, our work suggests that a fine control of the interplay between lattice 
vibrations and plasma oscillations may offer a pathway for investigating novel emergent 
states in quantum materials, and provide opportunities in the development of quantum technologies 
based on TMOs.

\vspace{10pt}

\section*{Methods} 

\section*{\small Ground-state calculations} \vspace*{-10pt}

{\it Ab initio} calculations were carried out for anatase TiO$_2$ (space group $I4_1/amd$), using
the experimental lattice parameter $a=3.784$~\AA\cite{Horn1972}. We used density-functional theory
(DFT) within the generalized gradient approximation of Perdew, Burke, and Ernzerhof\cite{PBE1996}.
The core-valence interaction was described by means of norm-conserving pseudopotentials, with the
semicore Ti-$3s$ and Ti-$3p$ states explicitly taken into account. Electron wavefunctions were
expanded in a planewaves basis set with kinetic energy cutoff of 200~Ry, and the Brillouin zone
was sampled using a $6\times6\times6$ Monkhorst-Pack mesh. Lattice-dynamical properties were
calculated using  density functional perturbation theory (DFPT). All DFT and DFPT calculations 
were performed using the Quantum ESPRESSO\cite{QuantumEspresso} package.

\section*{\small Electron-phonon coupling} \vspace*{-10pt}

Calculations of electron-phonon couplings were performed using the \texttt{EPW} code\cite{EPW2016},
the cumulant expansion was performed separately (see Supplementary Note~1). The Fr\"ohlich electron-phonon 
matrix element was calculated using the method of ref.~\citenum{Verdi2015}. An accurate description of 
the Fr\"ohlich vertex as described in ref.~\citenum{Verdi2015} is essential: ignoring this effect leads 
to a severe underestimation of the mass enhancement\cite{Zhukov2014}. To evaluate Eq.~\eqref{Sigma} 
we computed electronic and vibrational states as well as the scattering matrix elements on a 
$4\times4\times4$ Brillouin-zone grid. These quantities were interpolated with \textit{ab initio} 
accuracy onto a fine grid with $2\cdot10^6$ random $\bq$-points using \texttt{EPW}. The positive 
infinitesimal $\eta$ was set to 10~meV. Temperature effects were accounted for by including the 
Fermi-Dirac occupation in the spectral function, corresponding to the experimental temperature of 20~K.

\section*{\small Doping} \vspace*{-10pt}

Doping was included using the rigid-band approximation since the system is degenerate. 
In fact, by considering the Mott criterion for the metal-insulator transition\cite{Mott1990}, 
$a_0^\ast\,n_c^{1/3}\approx0.25$ where $a_0^\ast=\hbar^2\epsilon_0/(e^2 m_{\rm b})$ is the effective 
Bohr radius, we obtain a value $n_c=1.3\times10^{18}$~cm$^{-3}$ for the critical density that is below 
the doping levels investigated. The screening of the electron-phonon interaction arising from the 
doped carriers was taken into account by computing the dielectric function $\epsilon(\bq,\omega)$ in 
the random-phase approximation\cite{Mahan, Hedin1965}, for a homogeneous electron gas with the calculated 
effective mass $m_b$ and dielectric permittivity $\epsilon_\infty$ of anatase TiO$_2$\cite{Mahan, Hedin1965} 
(see Supplementary Note~3 for a rationale). In this expression the plasma frequency directly reflects 
the carrier density, therefore the influence of doping on the dielectric screening in anatase 
TiO$_2$\cite{Mardare2004} is included in the calculations. The resulting non-adiabatic matrix element 
is $g_{mn\nu}^{\rm NA}(\bk,\bq)= g_{mn\nu}(\bk,\bq)/\epsilon(\bq,\omega_{\bq\nu}+i/\tau_{n\bk})$, where 
$\tau_{n\bk}$ is the electron lifetime near the band edge. Here we use $\hbar/\tau_{n\bk} = 55$~meV 
from ref.~\citenum{Moser2013}. We checked that this choice does not affect our conclusions (Supplementary 
Fig.~4). We note that, owing to the strong anisotropy of the Fermi surface in \mbox{$n$-doped} anatase 
TiO$_2$, it is possible that the nominal doping levels estimated in ref.~\citenum{Moser2013} using the 
Fermi momentum may underestimate the actual carrier densities in the measured samples.

\section*{\textsf{\small Angle-resolved photoemission spectra}} \vspace*{-10pt}

The cumulant function used in the calculation of the ARPES spectra is obtained from the self-energy of 
Eq.~\eqref{Sigma} as follows\cite{Aryasetiawan1996}:
$ C_{n\bk}(t)=(\pi\hbar)^{-1}\int_{-\infty}^{\infty}{\rm d}\omega \,
 \,\theta(\omega-\ve_{n\bk}/\hbar)\, e^{(i\omega+\eta)t}\,
 {\rm Im}\,\Sigma_{n\bk}(\ve_{n\bk}/\hbar-\omega)/(\omega-i\eta)^2
$, where $\theta$ is the Heaviside function. More details on the cumulant expansion are provided 
in Supplementary Note~1. The calculations within the one-shot Migdal approximation
shown in Fig.~\ref{fig2}b were performed by using directly Eq.~\eqref{Sigma} inside the spectral 
function\cite{GiustinoRMP}: 
$ A(\bk,\omega)= 2\pi^{-1}\sum_n |\,{\rm Im}\,(\,\hbar\omega-\epsilon_{n\bk}-\Sigma_{n\bk})^{-1}| $.
The spectra were computed without including photon matrix elements effects\cite{Damascelli2003}. 
This approximation is justified since the variation of the orbital character over the $\sim$0.1~eV 
binding energy range shown in Fig.~\ref{fig1} is negligible\cite{Damascelli2003}. 
In order to extract the bands shown in Fig.~\ref{fig1}g-i we proceeded as follows. For the 
quasiparticle bands, which are the brightest features in Fig.~\ref{fig1}d-f, we calculate the poles 
of the electron Green's function as $E_{n\bk}=\ve_{n\bk}+{\rm Re}\,\Sigma_{n\bk}(E_{n\bk})$, where 
$\ve_{n\bk}$ and $E_{n\bk}$ are the bare and the renormalized dispersions, respectively. For the 
satellites in Fig.~\ref{fig1}d-e the previous procedure is not applicable since these features are 
not poles of the Green's function. In this case we directly identify the maxima in the energy 
dispersion curves\cite{Damascelli2003}, and obtain the pairs of parabolic bands which are visible 
in Fig.~\ref{fig1}g-h at binding energies below 0.1~eV. 

\section*{\small Polaron wavefunction} \vspace*{-10pt}

In order to calculate the wavefunction of the polaron we generalized the perturbation theory approach 
of ref.~\citenum{Mahan} to the case of multiple electron bands and phonon branches, as well 
as {\it ab initio} electron-phonon matrix elements. We found
$ \tilde{\psi}_{n\bk}(\mathbf r;\{\bm\tau_{\kappa p}\})\!=\!C \psi_{n\bk}({\bf r}) \times\left[1\!+\!
 \sum_{\nu}\int\!\frac{d\bq}{\Omega_{\rm BZ}}\,|g^{\rm NA}_{nn\nu}(\bk,\bq)|^2\right.$ 
 $\left.e^{i\bq\cdot\mathbf r}/(\ve_{n\bk}\!-\!\ve_{n\bk+\bq}\!-\!\hbar\omega_{\bq\nu})^2 \right]\!\ket{0_{\rm p}}$, 
 where $\psi_{n\bk}$ and $\tilde{\psi}_{n\bk}$ are the electron wavefunction without and with 
 electron-phonon interactions, respectively. $\{\bm\tau_{\kappa p}\}$ indicate the nuclear coordinates, 
$\ket{0_{\rm p}}$ is the phonon ground state and $C$ is a normalization constant. The details of the 
derivation are given in Supplementary Note~4. The wavefunctions in Fig.~\ref{fig2}f were 
obtained by setting $n\bk$ to the conduction band edge; the envelope functions were calculated by retaining 
only the term inside the square brackets in the above expression. A random grid with $2\cdot10^6$ points 
was used for the Brillouin-zone integration, and a small imaginary part $\eta=10$~meV was added to the 
denominator.

\vspace{0.5cm}
\section*{Acknowledgements} \vspace*{-10pt}
The authors gratefully acknowledge Marco~Grioni for granting permission to reproduce Fig.~1a-c, and 
M.~Grioni, B.~Gumhalter and P.~D.~C. King for fruitful discussions. 
This work was supported by the Leverhulme Trust (Grant RL-2012-001), the Graphene Flagship 
(Horizon 2020 Grant No. 696656 -- GrapheneCore1), and the UK Engineering and Physical 
Sciences Research Council (Grant No. EP/M020517/1). This work used the ARCHER UK 
National Supercomputing Service via the ’AMSEC’ Leadership project, and the Advanced Research 
Computing facility of the University of Oxford~\cite{ARC}. 

\vspace{0.5cm}
\section*{References}
\bibliographystyle{naturemag}
\bibliography{biblio-tio2}

\clearpage



\section*{Supplementary material}

# Origin of the crossover from polarons to Fermi liquids in transition metal oxides

# Supplementary Information

Carla Verdi, Fabio Caruso, Feliciano Giustino

*Department of Materials, University of Oxford, Parks Road, Oxford OX1 3PH, United Kingdom*

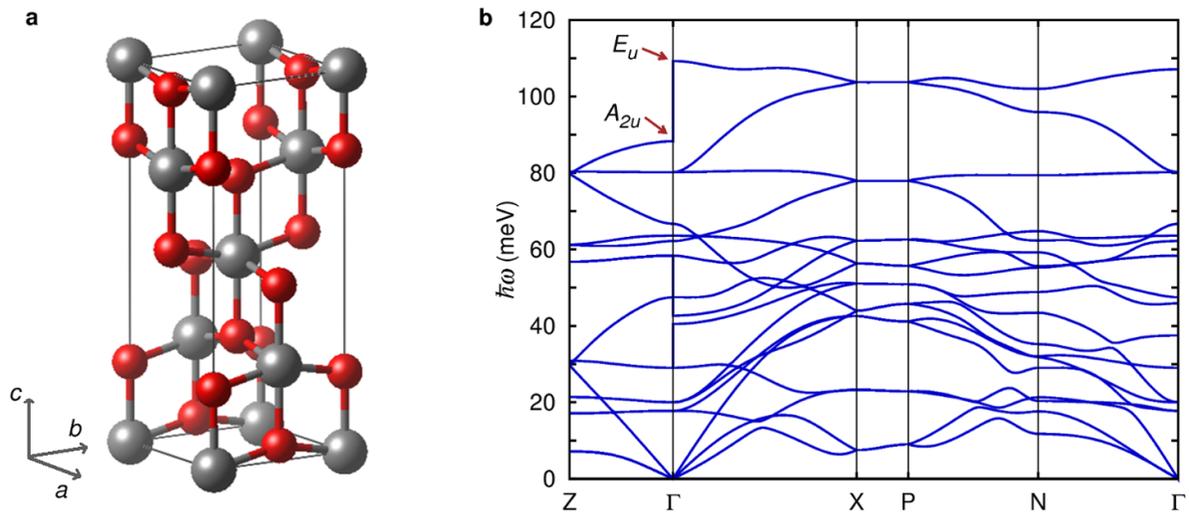

**Supplementary Figure 1 | Crystal structure and phonon dispersions of anatase TiO$_2$.**
(**a**) Ball-and-stick model of the tetragonal unit cell of anatase TiO$_2$. Ti atoms are in gray and O atoms are in red. The lattice vectors and crystallographic directions referred to in the main text are indicated. (**b**) Calculated phonon dispersion relations of anatase TiO$_2$ along high-symmetry directions of the Brillouin zone. The $E_u$ and $A_{2u}$ phonons discussed in the main text are indicated. The $A_{2u}$ phonon is infrared active only along the $c$ axis, accordingly it is found to give a smaller contribution to the total Fröhlich coupling after integrating over all phonon momenta.

2

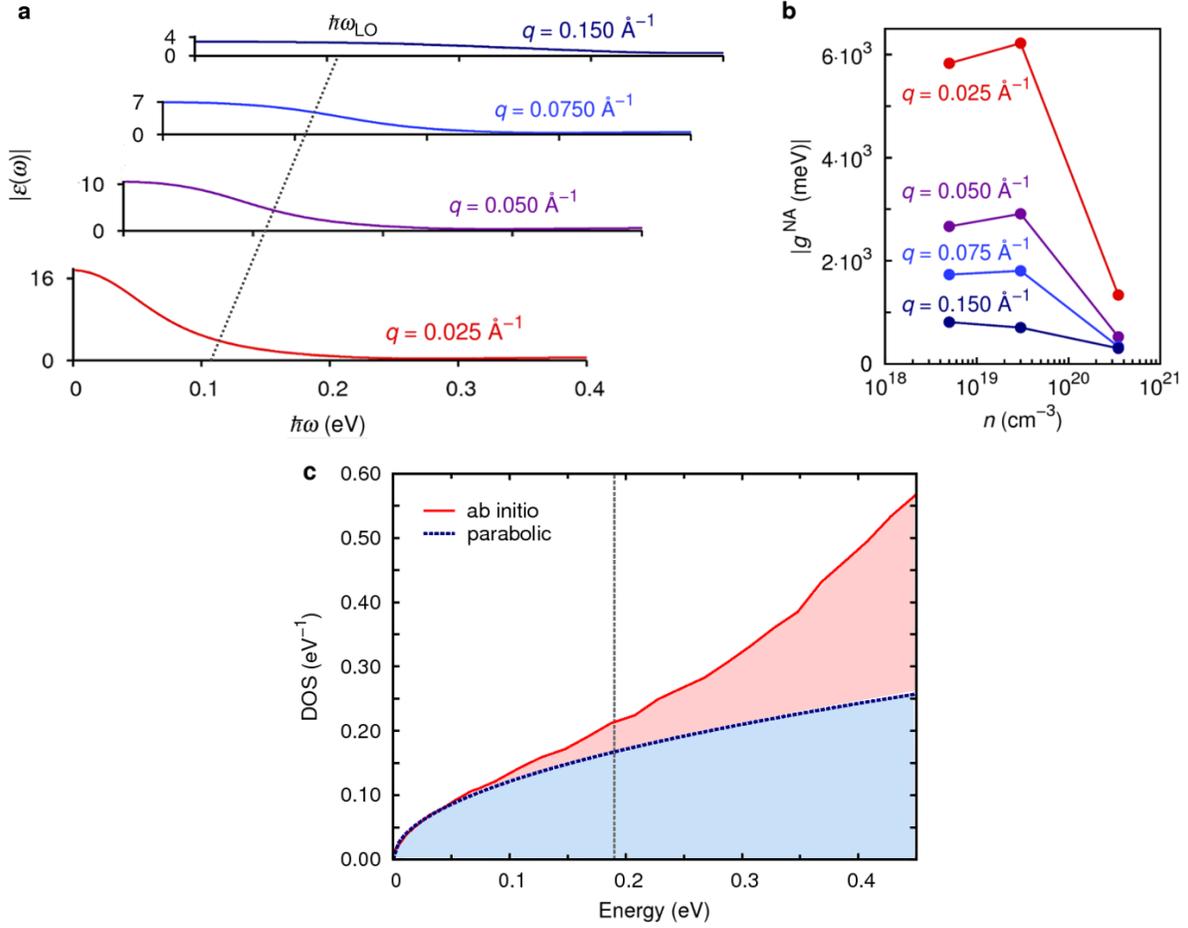

**Supplementary Figure 2 | Non-adiabatic electron-phonon matrix elements in anatase TiO$_2$.**
(**a**) Lindhard dielectric function associated with doped carriers in anatase TiO$_2$ at different values of **q**. For illustration we consider the highest doping level, $3.5 \times 10^{20}$ cm$^{-3}$. (**b**) Non-adiabatic electron-phonon matrix elements $g^{\mathrm{NA}}_{mn\nu}(\mathbf{k},\mathbf{q})$ corresponding to the LO E$_u$ phonon of anatase TiO$_2$, as a function of doping level. We set $n\mathbf{k}$ to the bottom of the conduction band. The electron-phonon matrix element becomes weaker at high doping. (**c**) Comparison between the density of states (DOS) near the conduction band bottom calculated from first principles (red) and modelled with a parabolic band (blue): $\mathrm{DOS}(E) = V/(2\pi^2)\,(2m_{\mathrm{b}}/\hbar^2)^{3/2}\sqrt{E}$, where $V$ is the unit cell volume of anatase. The *ab initio* DOS was computed using Wannier interpolation to obtain the electronic energies on a randomly generated dense **k** grid and with a Gaussian smearing of 5 meV. The gray vertical line indicates the energy corresponding to the Fermi level at the highest doping. The parabolic DOS starts deviating significantly from the calculated one for energies above 0.2 eV.

3

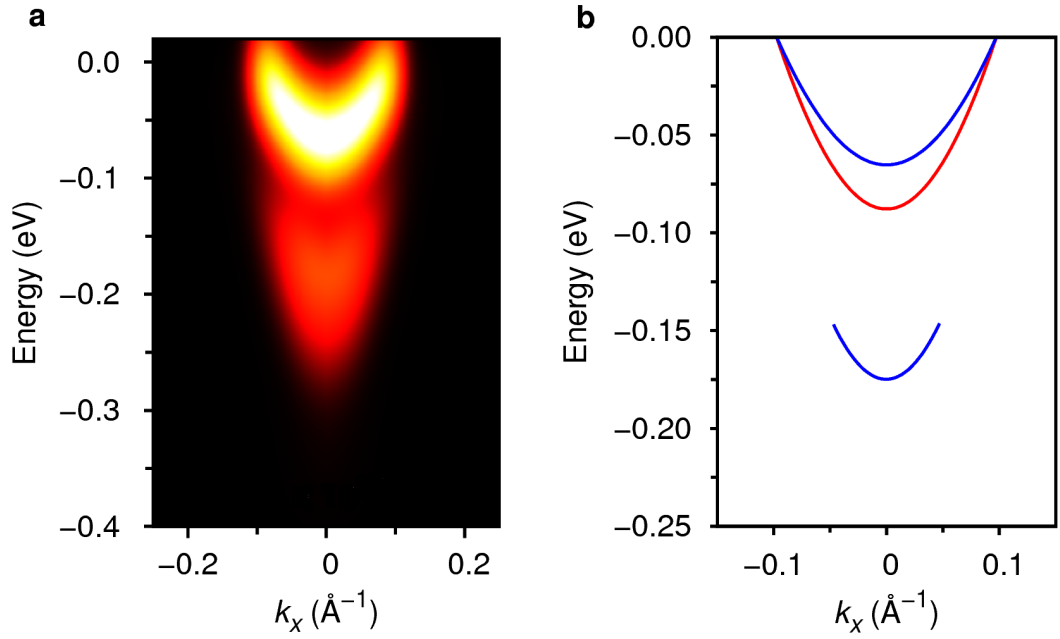

**Supplementary Figure 3 | Spectral function in the transition regime.** (**a**) Spectral function of anatase TiO$_2$ calculated for the doping level $1 \times 10^{20}$ cm$^{-3}$. Gaussian masks of widths 25 meV and 0.015 Å$^{-1}$ were applied as in Fig. 1 of the main text. (**b**) Band structure extracted from **a** (blue lines) together with the bare band (red line). For this doping one satellite is still visible but the mass renormalization parameter is decreasing ($\lambda = 0.34$ as reported in the main text).

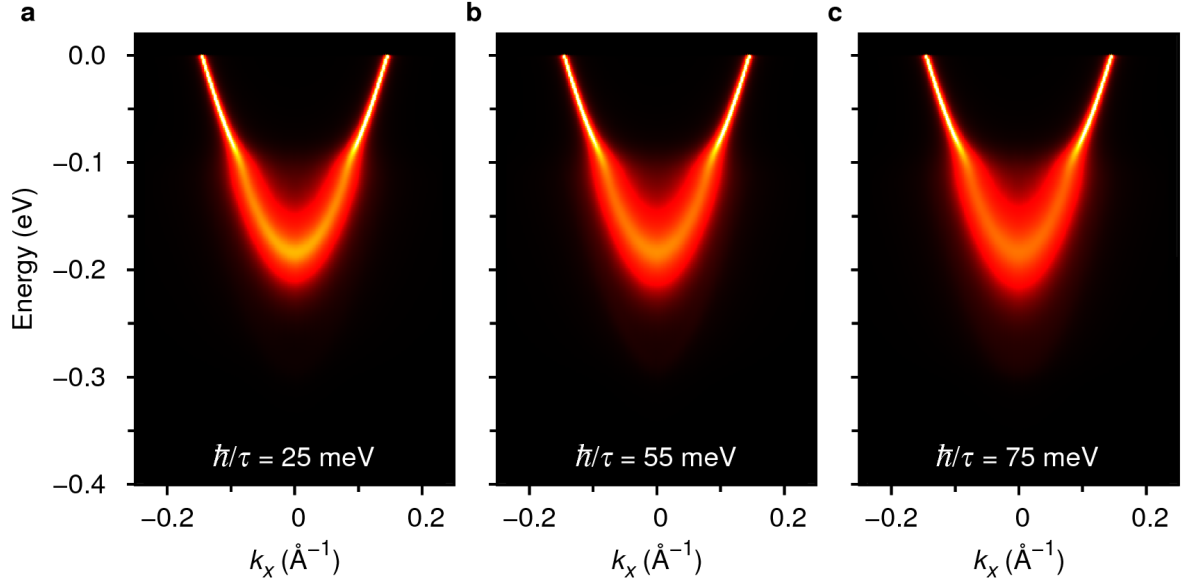

**Supplementary Figure 4 | Impact of the electron lifetime on the spectral properties.** Spectral function of anatase TiO$_2$ calculated for the doping level $3.5 \times 10^{20}$ cm$^{-3}$ using $\hbar/\tau = 25$ (**a**), 55 (**b**) and 75 meV (**c**) to compute the electronic screening (see Methods). The spectral features are virtually unchanged. We recomputed the mass renormalization parameter $\lambda$ as in the main text which gives $\lambda = 0.19$, 0.20 and 0.22 for increasing broadening. The same analysis for the doping concentration $5 \times 10^{18}$ cm$^{-3}$ yields $\lambda = 0.77$, 0.73 and 0.72 for the same broadenings. These values fall within 10% of the results presented in the main text.

**Supplementary Note 1 | Cumulant expansion**

The spectral function is related to the imaginary part of the one-electron retarded Green's function by:

$$A(\mathbf{k},\omega) = -\frac{1}{\pi}\sum_n \operatorname{Im} G_{n\mathbf{k}}(\omega). \tag{1}$$

In the cumulant expansion the Green's function is obtained in the time domain, starting from the the interaction picture[1-4]:

$$G_{n\mathbf{k}}(t) = i \, \exp\left[-i\varepsilon_{n\mathbf{k}}t/\hbar + C_{n\mathbf{k}}(t)\right], \tag{2}$$

where $C_{n\mathbf{k}}(t)$ is the cumulant function. By taking the Fourier transform of this expression to the frequency domain and inserting it in Supplementary Eq. (1) we obtain Eq. (1) of the main text. In order to obtain an expression for the cumulant which is amenable to computation, it is customary to expand the exponential in powers of the cumulant: $G_{n\mathbf{k}} = G^0_{n\mathbf{k}}\left[1 + C_{n\mathbf{k}} + C^2_{n\mathbf{k}}/2 + \cdots\right]$, where $G^0_{n\mathbf{k}}(t) = i \, \exp\left(-i\varepsilon_{n\mathbf{k}}t/\hbar\right)$ is the Green's function in absence of electron-phonon interactions. Alternatively, the Green's function can be obtained from the Dyson equation $G_{n\mathbf{k}} = G^0_{n\mathbf{k}} + G^0_{n\mathbf{k}}\Sigma_{n\mathbf{k}}G^0_{n\mathbf{k}} + G^0_{n\mathbf{k}}\Sigma_{n\mathbf{k}}G^0_{n\mathbf{k}}\Sigma_{n\mathbf{k}}G^0_{n\mathbf{k}} + \cdots$. By comparing these expansions term-by-term one finds an explicit expression for the cumulant function, which is given in the Methods and is reproduced here for completeness:

$$C_{n\mathbf{k}}(t) = \frac{1}{\pi\hbar}\int_{-\infty}^{\infty} d\omega \, \frac{\operatorname{Im}\Sigma_{n\mathbf{k}}(\varepsilon_{n\mathbf{k}}/\hbar - \omega)}{(\omega - i\eta)^2} e^{(i\omega+\eta)t} \, \theta(\varepsilon_{\mathrm{F}} - \varepsilon_{n\mathbf{k}} + \hbar\omega). \tag{3}$$

In practical calculations the self-energy $\Sigma_{n\mathbf{k}}$ is replaced by the best available approximation, which is the Migdal expression given in Eq. (2) of the main text. A rigorous derivation of the cumulant expansion and a discussion of its advantages and limitations can be found in Supplementary Refs. 4,5. In particular, in Supplementary Ref. 4 it is shown that the choice of seeding the self-energy calculated with the first non-crossing diagram also guarantees that no overcounting of correlated higher-order contributions is introduced in the theory. The inclusion of crossing diagrams in the evaluation of the Green's function results from the time orderings of the $t$ variables in the cumulant expansion.

The spectral function given in Eq. (1) of the main text yields multiple bosonic satellites, one for each term in the Taylor expansion of $\exp[C_{n\mathbf{k}}(t)]$. A convenient expression for the case of a

single satellite was derived in Supplementary Ref. 3. In this work we extended the method of Supplementary Ref. 3 to the case of multiple satellites. Following Supplementary Ref. 3, we write the spectral function as:

$$A(\mathbf{k},\omega) = \sum_n \left[ A^{\rm QP}_{n\mathbf{k}}(\omega) + A^{\rm QP}_{n\mathbf{k}}(\omega) * A^{\rm S1}_{n\mathbf{k}}(\omega) + A^{\rm QP}_{n\mathbf{k}}(\omega) * A^{\rm S2}_{n\mathbf{k}}(\omega) + \cdots \right], \quad (4)$$

where $*$ indicates the convolution. In the last expression the quasiparticle contribution $A^{\rm QP}_{n\mathbf{k}}(\omega)$ is defined as:

$$A^{\rm QP}_{n\mathbf{k}}(\omega) = \frac{2}{\pi} \frac{|{\rm Im}\Sigma_{n\mathbf{k}}(\varepsilon_{n\mathbf{k}})|}{[\hbar\omega - \varepsilon_{n\mathbf{k}} - {\rm Re}\Sigma_{n\mathbf{k}}(\varepsilon_{n\mathbf{k}})]^2 + [{\rm Im}\Sigma_{n\mathbf{k}}(\varepsilon_{n\mathbf{k}})]^2}, \quad (5)$$

and the satellite contributions associated with one-phonon and two-phonon processes are:

$$A^{\rm S1}_{n\mathbf{k}}(\omega) = \int_{-\infty}^{\infty} dt\, e^{i\omega t} C_{n\mathbf{k}}(t), \qquad A^{\rm S2}_{n\mathbf{k}}(\omega) = \int_{-\infty}^{\infty} dt\, e^{i\omega t} \frac{1}{2} C_{n\mathbf{k}}(t) C_{n\mathbf{k}}(t). \quad (6)$$

The function $A^{\rm S1}$ can be written in terms of the electron-phonon self-energy by combining Supplementary Eqs. (6) and (3) and carrying out the Fourier transform. This step was performed in Supplementary Ref. 3:

$$A^{\rm S1}_{n\mathbf{k}}(\omega) = \frac{\beta_{n\mathbf{k}}(\omega) - \beta_{n\mathbf{k}}(\varepsilon_{n\mathbf{k}}) - (\omega - \varepsilon_{n\mathbf{k}}/\hbar)\left.\frac{\partial \beta_{n\mathbf{k}}}{\partial \omega}\right|_{\varepsilon_{n\mathbf{k}}/\hbar}}{(\hbar\omega - \varepsilon_{n\mathbf{k}})^2}, \quad (7)$$

with

$$\beta_{n\mathbf{k}}(\omega) = \frac{1}{\pi} {\rm Im}\, \Sigma_{n\mathbf{k}}(\varepsilon_{n\mathbf{k}}/\hbar - \omega) \theta(\varepsilon_{\rm F}/\hbar - \omega). \quad (8)$$

In order to obtain the contribution of the second satellite, we note that $A^{\rm S1}_{n\mathbf{k}}(\omega)$ is the Fourier transform of $C_{n\mathbf{k}}(t)$, therefore the expression for $A^{\rm S2}_{n\mathbf{k}}(\omega)$ in Supplementary Eq. (6) can be rewritten using the convolution theorem:

$$A^{\rm S2}_{n\mathbf{k}} = \frac{1}{2} A^{\rm S1}_{n\mathbf{k}} * A^{\rm S1}_{n\mathbf{k}}. \quad (9)$$

By combining Supplementary Eqs. (9) and (4) we obtain the expression used in our calculations:

$$A(\mathbf{k},\omega) = \sum_n \left[ A^{\rm QP}_{n\mathbf{k}}(\omega) + A^{\rm S1}_{n\mathbf{k}}(\omega) * A^{\rm QP}_{n\mathbf{k}}(\omega) + \frac{1}{2} A^{\rm S1}_{n\mathbf{k}}(\omega) * A^{\rm S1}_{n\mathbf{k}}(\omega) * A^{\rm QP}_{n\mathbf{k}}(\omega) + \cdots \right]. \quad (10)$$

From Supplementary Eq. (10) we can extrapolate the general expression for the case of many satellites:

$$A(\mathbf{k},\omega) = \sum_n \left[ 1 + A^{\rm S1}_{n\mathbf{k}}(\omega) * + \frac{1}{2} A^{\rm S1}_{n\mathbf{k}}(\omega) * A^{\rm S1}_{n\mathbf{k}}(\omega) * + \cdots \right] A^{\rm QP}_{n\mathbf{k}}(\omega) = \sum_n e^{A^{\rm S1}_{n\mathbf{k}}(\omega)*} A^{\rm QP}_{n\mathbf{k}}(\omega). \quad (11)$$

7

When considering a drastically simplified model system consisting of dispersionless electrons and an Einstein phonon spectrum, this last expression reduces to the well-known Lang-Firsov series of polaron satellites[1,6]. In fact, by setting $A^{\mathrm{QP}} = Z\delta(\omega)$ and $A^{\mathrm{S1}} = \lambda\delta(\omega - \omega_{\mathrm{ph}})$, Supplementary Eq. (11) gives $A(\omega) = Z\sum_{m=0}^{\infty} (\lambda^m/m!)\,\delta(\omega - m\,\omega_{\mathrm{ph}})$. By further requiring the normalization of the spectral function, that is $\int A(\omega)\,\mathrm{d}\omega = 1$, we obtain $Z = e^{-\lambda}$ and the Lang-Firsov expression is recovered. In our *ab initio* calculations we truncate Supplementary Eq. (11) to the second order. In practice we first evaluate the quasiparticle and satellite contributions, and then we perform two successive numerical convolutions in order to obtain the two satellites which are seen in the experiments.

**Supplementary Note 2 | Mass enhancement and Fröhlich coupling constant**

The mass enhancement parameters obtained from the quasiparticle bands along the $\Gamma\Sigma$, $\Gamma$X, and $\Gamma$Z directions are $\lambda = 0.73$, 0.73, and 0.74 at the lowest doping; $\lambda = 0.70$ in each direction at the intermediate doping; and $\lambda = 0.20$, 0.20, and 0.18 at the highest doping. The mass enhancement can also be calculated directly from the energy derivative of the real part of the electron self-energy at the Fermi level, $\lambda_{\mathbf{k}} = -\partial\mathrm{Re}\Sigma_{\mathbf{k}}(\omega)/\partial\omega|_{\omega=\varepsilon_{\mathrm{F}}}$[7]. We checked that the values thus obtained fall within less than 10% of the ones listed above, the average over the Fermi surface being $\lambda = 0.68, 0.65$ and $0.19$ for the three values of doping considered. This approach also validates our calculations of the contribution to $\lambda$ arising from different phonon modes presented in the analysis of Fig. 2a in the main text.

From the mass enhancement parameter it is also possible to obtain the quasiparticle strength $Z$ as $Z = 1/(1 + \lambda)$ [7]. Using the values of $\lambda$ reported in Fig. 1g-i we obtain $Z = 0.58$, 0.59, and 0.83 with increasing doping. Our calculated quasiparticle strength at intermediate doping overestimates the experimentally-determined value, $Z = 0.36$ [8]. We attribute this difference to extrinsic losses not accounted for in our calculations; these losses are known to transfer spectral weight to the satellites[3,9,10].

In polar materials it is customary to describe the coupling to a dispersionless LO phonon via the

8

dimensionless Fröhlich coupling constant $\alpha$:

$$\alpha = \frac{e^2}{\hbar}\left(\frac{m_{\rm b}}{2\hbar\omega_{\rm LO}}\right)^{1/2}\left(\frac{1}{\epsilon_\infty}-\frac{1}{\epsilon_0}\right), \tag{12}$$

where $m_{\rm b}$ is the band mass, $\hbar\omega_{\rm LO}$ the energy of the LO phonon, $\epsilon_0$ and $\epsilon_\infty$ the static and high-frequency permittivities, respectively. We evaluate Supplementary Eq. (12) using the experimental values of $\omega_{\rm LO}$, $\epsilon_0$, and $\epsilon_\infty$ from Supplementary Ref. 11. For the effective masses we use our DFT calculations since no accurate experimental values are available: $m_\perp = 0.40\,m_{\rm e}$ and $m_\parallel = 4.03\,m_{\rm e}$. By using these parameters in Supplementary Eq. (12) we find the experimental Fröhlich constants $\alpha_\perp = 1.0$ and $\alpha_\parallel = 3.4$, and their isotropic average $\alpha = 1.8$. The Fröhlich constant $\alpha$ can also be obtained from *ab initio* calculations of the electron-phonon matrix elements, following Supplementary Ref. 12. In this case we calculate $\alpha_\perp = 0.9$, $\alpha_\parallel = 3.0$, and the isotropic average $\alpha = 1.6$. These values are in excellent agreement with experiment, therefore our description of Fröhlich coupling in anatase $TiO_2$ is expected to be highly accurate. We note that the anisotropy of the Fröhlich coupling constant $\alpha_\perp$ and $\alpha_\parallel$ does not stem from anisotropic electron-phonon interactions, but rather from the strong anisotropy in the band masses, which enter $\alpha$ as seen in Supplementary Eq. (12).

For weak and intermediate couplings, the polaron effective mass is commonly estimated using $m^* = m_{\rm b}(1-\alpha/6)^{-1}$ [13,14]. While this procedure is adequate for isotropic crystals, it cannot be used in the present case of $TiO_2$. In fact, if we use the isotropic average of the coupling constant, then we underestimate $m^*$ with respect to experiment[8]. On the other hand, if we consider $\alpha_\perp$ and $\alpha_\parallel$ separately, then we obtain a strongly anisotropic mass enhancement, which is not consistent with our many-body calculations. These observations indicate that, in the case of anisotropic crystals, the Fröhlich constant $\alpha$ should be used with caution.

**Supplementary Note 3 | Dielectric screening**

In this work we calculate the additional screening arising from the charge carriers using the random phase approximation (RPA) for the homogeneous electron gas, that is the Lindhard screening. This choice is motivated by the fact that we must evaluate the screening for millions of electron-phonon matrix elements, and explicit *ab initio* calculations of the RPA screening for such a dense Brillouin

9

zone grid are not currently feasible. The use of the Lindhard model is justified by the fact that the system under consideration lies in the high-density electron-gas limit. To confirm this point we evaluate the Wigner-Seitz radius $r_s$ given by $r_s = (4\pi n a_0^{*3}/3)^{-1/3}$, where $n$ is the doping density and $a_0^* = \hbar^2 \epsilon_0/(e^2 m_{\rm b})$[15]. Using the band effective mass $m_{\rm b}$ and the static dielectric constant $\epsilon_0$ of anatase we obtain $r_s = 1.6$ for the lowest doping level $5 \times 10^{18}$ cm$^{-3}$. This value is comparable or even smaller than what found in simple metals, therefore the use of the electron gas model to describe doped carriers is justified. Moreover, the electronic bands are to a good approximation parabolic in the energy range considered in this work; this is seen by comparing the *ab initio* density of states with the parabolic band model, Supplementary Fig. 2c. Therefore the use of a Lindhard function is justified, and we expect this choice to be very accurate in the present case.

**Supplementary Note 4 | Polaron wavefunction**

In order to calculate the wavefunction of a polaron, we follow the approach of Supplementary Ref. 15 and express the many-body electron-phonon state using Rayleigh-Schrödinger perturbation theory. We consider a system at zero temperature and with a single electron added to the bottom of the conduction band. The resulting expression is:

$$\tilde{\psi}_{n\mathbf{k}}(\mathbf{r};\{\boldsymbol{\tau}_\kappa\}) = \psi_{n\mathbf{k}}(\mathbf{r})\left|0_{\rm p}\right\rangle + \frac{1}{\sqrt{N_{\mathbf{q}}}}\sum_{m\nu\mathbf{q}}\frac{g_{mn\nu}(\mathbf{k},\mathbf{q})\psi_{m\mathbf{k}+\mathbf{q}}(\mathbf{r})}{\varepsilon_{n\mathbf{k}} - \varepsilon_{m\mathbf{k}+\mathbf{q}} - \omega_{\mathbf{q}\nu}}\,\hat{a}^\dagger_{-\mathbf{q}\nu}\left|0_{\rm p}\right\rangle, \quad (13)$$

where $\{\boldsymbol{\tau}_\kappa\}$ are the nuclear coordinates, $\hat{a}^\dagger_{-\mathbf{q}\nu}$ is a phonon creation operator, and $\left|0_{\rm p}\right\rangle$ is the ground state with no phonons. The Brillouin zone is discretized using a uniform grid with $N_{\mathbf{q}}$ phonon wavevectors. Since the atomic displacements are smaller than characteristic interatomic distances, we can simplify the above expression by replacing $\hat{a}^\dagger_{-\mathbf{q}\nu}$ (which is a function of the normal mode coordinates) by its average over a given electron-phonon state[15]. In order to determine a lower bound to the polaron radius, we evaluate this expectation value by considering an electronic wavefunction localized at the center of the reference frame, $\mathbf{r} = 0$. The result is:

$$\langle \hat{a}^\dagger_{-\mathbf{q}\nu}\rangle = \frac{1}{\sqrt{N_{\mathbf{q}}}}\sum_{m'}\frac{g^*_{m'n\nu}(\mathbf{k},\mathbf{q})}{\varepsilon_{n\mathbf{k}} - \varepsilon_{m'\mathbf{k}+\mathbf{q}} - \omega_{\mathbf{q}\nu}}. \quad (14)$$

By replacing $\langle \hat{a}^\dagger_{-\mathbf{q}\nu}\rangle$ inside Supplementary Eq. (13) we have:

$$\tilde{\psi}_{n\mathbf{k}}(\mathbf{r};\{\boldsymbol{\tau}_\kappa\}) = \frac{e^{i\mathbf{k}\cdot\mathbf{r}}}{\sqrt{\Omega}}\left[u_{n\mathbf{k}}(\mathbf{r}) + \frac{1}{N_\mathbf{q}}\sum_{m\nu\mathbf{q}}\frac{g_{mn\nu}(\mathbf{k},\mathbf{q})u_{m\mathbf{k}+\mathbf{q}}(\mathbf{r})e^{i\mathbf{q}\cdot\mathbf{r}}}{\varepsilon_{n\mathbf{k}}-\varepsilon_{m\mathbf{k}+\mathbf{q}}-\omega_{\mathbf{q}\nu}}\sum_{m'}\frac{g^*_{m'n\nu}(\mathbf{k},\mathbf{q})}{\varepsilon_{n\mathbf{k}}-\varepsilon_{m'\mathbf{k}+\mathbf{q}}-\omega_{\mathbf{q}\nu}}\right]|0_\mathrm{p}\rangle.$$
(15)

The expression for $\tilde{\psi}_{n\mathbf{k}}$ given in the Methods was obtained from Supplementary Eq. (15) by considering the normalization of the wavefunction and by retaining the long-wavelength part of the Fröhlich vertex, so that $g_{mn\nu}(\mathbf{k},\mathbf{q}) \to g_{mn\nu}(\mathbf{k},\mathbf{q})\,\delta_{mn}$. We already demonstrated that this approximation is very accurate for anatase TiO$_2$[12]. Since the main contribution to Supplementary Eq. (13) arises from long-wavelength optical phonons, we also replace $u_{m\mathbf{k}+\mathbf{q}}(\mathbf{r})$ by $u_{m\mathbf{k}}(\mathbf{r})$, in the spirit of the $\mathbf{k}\cdot\mathbf{p}$ approximation.

Our choice of calculating the polaron wavefunction within Rayleigh-Schrödinger perturbation theory is justified since this theory is valid in the range $\alpha \leq 5$ (see Supplementary Ref. 15, p. 516), and the largest Fröhlich coupling constant in anatase TiO$_2$ is $\alpha_\parallel = 3.0$ (see Supplementary Note 2).

We note that the present many-body approach for calculating the polaron wavefunction differs significantly from DFT calculations of excess electrons in oxides using large supercells[16–19]. In fact in the present case we employ a quantum description of both electrons and nuclei, and a dynamical (non-adiabatic) description of their interactions. In contrast, supercell DFT calculations describe the nuclei as classical particles and decouple electronic and nuclear degrees of freedom using the Born-Oppenheimer adiabatic approximation. These approximations are not justified in the case of TiO$_2$, and have led to results which are very sensitive to the DFT exchange and correlation functional and the size of the supercell. Our present perturbative approach does not suffer from these shortcomings.

## Supplementary References



\end{document}